
\tolerance=10000
\raggedbottom

\baselineskip=15pt
\parskip=1\jot

\def\sk{\vskip 3\jot}

\def\heading#1{\vskip3\jot{\noindent\bf #1}}
\def\label#1{{\noindent\it #1}}
\def\QED{\hbox{\rlap{$\sqcap$}$\sqcup$}}


\def\ref#1;#2;#3;#4;#5.{\item{[#1]} #2,#3,{\it #4},#5.}
\def\refinbook#1;#2;#3;#4;#5;#6.{\item{[#1]} #2, #3, #4, {\it #5},#6.} 
\def\refbook#1;#2;#3;#4.{\item{[#1]} #2,{\it #3},#4.}


\def\({\bigl(}
\def\){\bigr)}




\def\[{\big[}
\def\]{\big]}

\def\F{Fibonacci}
\def\Z{Zeckendorf}
\def\repr{representation}
\def\Zr{\Z{} \repr}

\def\u{\overline{1}}

{
\pageno=0
\nopagenumbers
\rightline{\tt eaza.tex}
\vskip1in

\centerline{\bf Efficient Algorithms for \Z{} Arithmetic}
\vskip0.5in

\centerline{Connor Ahlbach}
\centerline{\tt Connor\_Ahlbach@hmc.edu}
\sk

\centerline{Jeremy Usatine}
\centerline{\tt Jeremy\_Usatine@hmc.edu}
\sk

\centerline{Nicholas Pippenger}
\centerline{\tt Nicholas\_Pippenger@hmc.edu}
\sk

\centerline{Department of Mathematics}
\centerline{Harvey Mudd College}
\centerline{301 Platt Boulevard}
\centerline{Claremont, CA 91711}
\vskip0.5in

\noindent{\bf Abstract:}
We study the problem of addition and subtraction using the \Z{} \repr{} of integers.
We show that both operations can be performed in linear time; in fact they can be performed by combinational logic networks with linear size and logarithmic depth.
The implications of these results for multiplication, division and square-root extraction are also discussed.
\vfill\eject
}

\heading{1. Introduction}

\Z{}  [Z] observed that every integer $X\ge 0$  can be represented as the sum of a unique subset of the \F{} numbers $\{F_n: n\ge 2\}$
in which no two consecutive \F{} numbers appear.
That is, we may write
$$X = \sum_{k\ge 2} x_k \, F_k$$ 
for a unique sequence $x_k$ such that $x_k\in\{0,1\}$ and $x_k x_{k+1} = 0$ for all $k\ge 2$.
This representation is called the {\it\Z\/} \repr{} (or sometimes the {\it
\F\/} \repr{}) of $X$.
\Z{} \repr{}s have applications in coding; they have been proposed for use in self-delimiting codes
by Apostolico and Fraenkel [A] and by Fraenkel and Klein [F2].
They can also be used for run-length-limited binary codes (in which neither $0$ nor $1$ may appear more than twice in succession).
In both of these applications, the main step of encoding is to convert an integer from its binary \repr{} to its 
\Z{} \repr{} (with the reverse conversion occurring at the receiving end).
Arithmetic involving integers in their \Z{} \repr{}s also occurs in an algorithm for playing Wythoff's game [W3] due to Silber [S].

The main problem we address in this paper is: given the \Z{} \repr{}s of two integers,
$X$ and $Y$, how can we find the \Z{} \repr{} of their sum $Z=X+Y$?
\Z{} addition, and other arithmetic operations in \Zr, have been discussed by Ligomenides and Newcomb [L], by Freitag and Phillips [F3, F4], and by Fenwick [F1], but none of these authors explicitly discuss  the resources required by the methods proposed.
Tee [T] reviews these methods and gives explicit bounds, but his bounds are extremely weak: his algorithm for \Z{} addition of two 
$n$-digit numbers runs in time $O(n^3)$.
In Section 2, we shall show that \Z{} addition of $n$-digit numbers can be performed  in  time $O(n)$; in fact, it can be carried out in three linear passes (in alternating directions) over the input sequence.
This algorithm lends itself to efficient parallel as well as serial implementation; in fact, it can be
carried out by combinatorial logic networks (acyclic interconnections of gates with bounded fan-in) having size (number of gates, used as an estimate of cost) $O(n)$ and simultaneously having depth (number of gates on the longest path from an input to an output, used as an estimate of delay, or parallel execution time) $O(\log n)$.
Apart from the constants implicit in the $O$-notation, this result is the best possible (because the computation
of any function (such as the most significant bit of the output) that depends on all the inputs digits requires at least linear size and at least logarithmic depth).
See Wegener [W2] for an excellent treatment of combinational logic networks as a computational model well suited to the discussion of resource bounds.

In Section 3, we shall extend our results on addition to subtraction (that is, to the addition of signed integers, represented by adjoining a positive or negative sign to their \Z{} \repr{}s).
Our result is that signed integers can also be added using three alternating passes over the input, so that all the bounds derived in the preceding section continue to hold, albeit with larger constants in the $O$-terms.

Finally, in Section 4, we discuss the implications of the foregoing results to the problems of multiplication, division (with remainder), and square-root extraction (with remainder).
For all three problems, there are combinational logic networks treating $n$-digit numbers with size $O(n^2)$ and depth
$O\((\log n)^2\)$.
Our algorithms for these problems are based on conversion from \Z{} to binary \repr{} (for which we describe networks of size $O(n^2)$ and depth $O(\log n)$), and from binary to \Zr{} (for which we describe networks of size $O(n^2)$ and depth $O\((\log n)^2\)$).
\vfill\eject

\heading{2. Addition}

In this section, we shall show how finite automata making three passes over the input can perform addition of \Z{} \repr{}s.
We assume that the problem is presented as a sequence of $0$'s, $1$'s and $2$'s, obtained from the \Z{} \repr{}s of the numbers to be added by adding the digits in each position independently.
This sequence will in general not be the \Z{} \repr{} of the sum, since (1) it may have consecutive $1$'s, and (2) it may contain $2$'s.
But we note that any $2$'s in this sequence are both immediately preceded and immediately followed by $0$'s (because the $1$'s in the numbers to be added that produced these $2$'s could not have other $1$'s in consecutive positions).

We shall compute the \Zr{} of the sum from this sequence in two stages.
In the first stage we shall eliminate the $2$'s, obtaining an intermediate sequence that contains only $0$'s and $1$'s, but that may contain consecutive $1$'s.
In the second stage we shall convert this intermediate sequence into the \Zr{} of the sum.

The first stage will be performed in one left-to-right pass over the input sequence.
(We assume that the most-significant digits are on the left and the least-significant are on the right, as in conventional radix \repr{}s.
We also assume that the sequence begins with a $0$, which may be appended if necessary with out changing the value represented.)
We describe the actions taken by the algorithm in terms of a ``moving window'' four positions wide.
The window begins at the leftmost four positions.
At each step it may change the values of the digits at positions within the window, in accordance with rules presented below.
After each step the window is shifted one position to the right.
When the window has reached the rightmost four positions, and any changes applicable to those positions have been made, a final ``cleanup'' operation will be performed.

The rules describing the changes to be made at each window position are as follows.
$$\eqalign{
020x &\mapsto 100x' \cr
030x &\mapsto 110x' \cr
021x &\mapsto 110x \cr
012x &\mapsto 101x \cr
}$$
\sk
Here $x$ denotes one of the symbols $0$, $1$ or $2$, and $x'$ denotes its successor:
$1$, $2$ or $3$.
These rules employ the symbol $3$ in addition to $0$, $1$ and $2$, but at the end of the first stage
all $3$'s, as well as all $2$'s, will have been eliminated.
If none of these rules are applicable, the symbols within the window are not changed.

The soundness of these rules (that is, the fact that they leave the value represented unchanged) follows immediately from the recurrence for the Fibonacci numbers.
It remains to prove their effectiveness (that is, that all $2$'s and $3$'s are eventually eliminated).
This effectiveness is a consequence of the following two facts.

\label{Fact 1:} Whenever a $3$ is created (which happens when a $2$ is incremented in the fourth position of the window), it is both preceded by a $0$ and followed by a $0$ when it reaches the second position of the window, and is thus eliminated at that time.

\label{Fact 2:} Every $2$ (whether present in the input sequence or created by incrementing a $1$ in the fourth position) is either (1) preceded by a $0$ and followed by a $0$ or a $1$ when it reaches the second position, and is thus eliminated at that time, or (2) is preceded by the sequence $01$ when it reaches the third position, and is thus eliminated at that time.

\label{Proof of Fact 1:}
A $3$ is created by incrementing a $2$ in the fourth position.
Any symbol that is incremented must be preceded by a $0$.
A $2$ that is incremented must have been present in the input, where it must have been followed by a $0$, and this $0$ cannot be incremented because it is not preceded by another $0$.
\QED

\label{Proof of Fact 2:}
Consider first a $2$ that is created by incrementing a $1$ in the fourth position.
Any symbol that is incremented must be preceded by a $0$.
A $1$ that is incremented must have been present in the input, where it must have been followed by either a $0$ or a $1$, and this $0$ or $1$ cannot be incremented because it is not preceded by another $0$.
Thus created $2$'s are eliminated when they reach the second position.

Consider then a $2$ that is present in the input, where it is both preceded and followed by $0$.
If the preceding $0$ is not incremented when it is in the fourth position, then the reasoning above applies.
If the preceding $0$ is incremented, then it must have been preceded by another $0$,
so that the sequence $01$ then precedes the $2$.
The $0$ that initially followed the $2$ cannot be incremented because it is not preceded by another $0$.
Thus this $2$ will be eliminated when it reached the third position.
\QED

After any applicable changes have been made with the window in its rightmost position,
there may still be a $2$ or $3$ in the third or fourth position of the window.
These may be cleaned up as follows.
If there is a $3$ in the third position, then it must be preceded and followed by $0$'s.
Then $030$ can be changed to $111$ without changing the value represented.
If there is a $2$ in the third position it must be preceded either by a $0$ or by the subsequence $01$, and it must be followed by a $0$.
Then $020$ can be changed to $101$, or $0120$ can be changed to $1010$ without changing the value represented.
If there is a $3$ in the fourth position, then it must be preceded  by a $0$.
Then $03$ can be changed to $11$ without changing the value represented.
If there is a $2$ in the fourth position it must be preceded either by a $0$ or by the subsequence $01$.
Then $02$ can be changed to $10$, or $012$ can be changed to $101$ without changing the value represented.
After this cleanup operation, the resulting intermediate sequence consists entirely of $0$'s and $1$'s.

The second stage will be performed in two passes over the intermediate sequence, the first from right to left, and the second from left to right.
The first, right-to-left pass will use a window of width three, and will make changes according to the following 
single rule.
$$ 011 \mapsto 100$$
The soundness of this rule is again follows immediately from the recurrence for the Fibonacci numbers.
After this right-to-left pass, the resulting sequence contains no occurrence of the subsequence $1011$.
Suppose, to obtain a contradiction, that the resulting sequence does contain an occurrence of the subsequence $1011$.
Focus on one such occurrence, and suppose that at step $s$ the window was positioned at the final three positions of this occurrence.
At the outset of step $s$, the final two positions of the occurrence must have been $1$, because at and after step $s$ they could only be changed from $1$'s to $0$s, not from $0$ to $1$'s.
Since they were not changed to $0$'s at step $s$, the second position of the occurrence must have been $1$ at step $s$, and thus must have subsequently been changed to $0$.
But there are only two steps that could have made this change: steps $s+1$ and $s+2$.
If step $s+1$ had made the change, it would also have put a $0$ in the third position of the occurrence, which would have remained forever after, contrary to what we see now.
And if step $s+2$ had made the change, it would also have put a $0$ in the first position of the occurrence, which would have remained forever after, contrary to what we see now.
Thus we have reached a contradiction, proving there is no occurrence of the subsequence $1011$ after the right-to-left pass.

The second pass of the second stage is a left-to-right pass using the same width-three window and the same rule.
It is easy to see that during this pass every pair of consecutive $1$'s is eliminated, no new pairs of consecutive $1$'s are created, and no occurrence of $1011$ is created.
Thus at the conclusion of stage two, the resulting sequence is the \Zr{} of the sum.

Since the operations performed in each pass can be carried out by a finite automaton making a single pass over the input, they can also be carried out by combinational logic networks 
having size $O(n)$ and depth $O(\log n)$,
where $n$ is the length of the input sequence.
\sk

\heading{3. Signed Numbers}

In this section, we shall extend the results of the preceding section to cover signed integers.
We shall assume that a sign bit ($0$ indicates non-negative, $1$ indicates non-positive) is appended to the \Zr.
Thus subtraction can be performed by flipping the sign of the number to be subtracted, then performing addition of signed numbers.

Consider the addition of two signed numbers.
If the numbers have the same sign, one adds the magnitudes of the numbers and appends the common sign to the sum.
This can be accomplished by trivial extension of the algorithm presented for unsigned addition.
If, on the other hand, the numbers have opposite signs, we must subtract the smaller magnitude from the larger magnitude, and append the sign of the number with larger magnitude.
There are two problems here: one is to determine which number has the greater magnitude,
and the other is to determine the difference between the magnitudes.

We shall describe the solution to the second problem first, assuming that we know that the positive number has greater magnitude.
After presenting this solution, we shall indicate how to modify it to solve the first problem as well.

We shall assume that the input is represented as a sequence of $0$'s, $+1$'s and $-1$'s, obtained from the \Zr{}s of the two numbers by combining the digits in each position separately, canceling 
$+1$'s and $-1$'s that occur in the same position to $0$'s.
For typographical convenience, we shall write $+1$'s and $-1$'s as $1$'s and $\u$'s, respectively.

Our algorithm will begin with a preliminary left-to-right pass over the input sequence.
The output from this pass will be a sequence of $0$'s $1$'s and $2$'s that can be used as input to the algorithm described in the preceding section to compute the difference between the magnitudes.
(Since the latter algorithm also begins with a left-to-right-pass, these two passes can be combined into a single pass with a wider window, so the algorithm for signed addition can also be performed in three alternating passes.)
The preliminary left-to-right pass will use a window of width three and apply the following rules.
$$\eqalign{
100 &\mapsto 011 \cr
1\u 0&\mapsto 001 \cr
1\u 1 &\mapsto 002 \cr
10\u &\mapsto 010 \cr
\cr
200 &\mapsto 111 \cr
2\u 0&\mapsto 101 \cr
2\u 1 &\mapsto 102 \cr
20\u &\mapsto 110 \cr
}$$
The strategy of these rules is clear.
They keep a symbol with positive sign in the window at all times, and use it to cancel symbols with negative sign whenever necessary.
They may need to introduce $2$'s, but any $2$'s in the output sequence are both preceded and followed by $0$'s.
Thus the output sequence satisfies the conditions required for the application of the addition algorithm in the preceding section, which can then be used to determine the difference between the magnitudes.

We must now address the problem of determining which of the signed numbers to be added has the greater magnitude.
To do this, we look for the first occurrence of a $1$ or $\u$ as we scan the input sequence from left-to-right; the sign of this occurrence will be the sign of the sum.
(If all the symbols of the input sequence are $0$, the sum is zero.)
One way to implement this process is to extend the window of the preliminary pass on the right  to a fourth position, which will then be the first to see the leading $1$ or $\u$.
(We assume that three $0$'s are prefixed to the input sequence so that the fourth position will be visited by every symbol of the original input.)
If it detects a leading $\u$ (rather than $1$), it can then flip the sign of that and every following symbol, and the result of the pass will then be the difference between the magnitudes, to which the sign of the leading symbol can be appended to give the final representation of the sum.
\sk

\heading{4. Multiplication, Division and Square-Root Extraction}

In this section, we shall show how our results on \Z{} addition can be used to give efficient algorithms for more complicated arithmetic operations.
We begin by observing that if we use our addition algorithm in Fenwick's [F1] algorithm for multiplication, we obtain combinational logic networks for \Z{} multiplication with size $O(n^2)$.
Unfortunately, these networks have depth $O(n\log n)$ (because $n$ additions, each requiring depth $(\log n)$, are performed serially).
While we know of no improvement to this size bound, we can improve the depth bound considerably, to $O\((\log n)^2\)$.

The strategy for the remaining results in this section is based on what Freitag and Phillips [F4] call ``cut[ting] the Gordian knot''.
We convert the inputs from \Zr{} to binary, perform the operation using a known efficient binary algorithm, and convert the result back to \Zr.
To implement this strategy, we must look at algorithms for converting between \Z{} and binary 
\repr{}s, and review some known results on binary arithmetic.

Our first observation is that a an $n$-bit number in binary can be converted to \Zr{} by a combinational logic network of size $O(n^2)$ and depth $O\((\log n)^2\)$.
To see this, we observe that we can construct a tree of \Z{} adders that computes the sum of $n$ $n$-digit numbers with size  
$O(n^2)$ and depth $O\((\log n)^2\)$.
(The tree contains $O(n)$ adders, each of size $O(n)$ and depth $O(\log n)$, each constructed according to the result presented in Section 2.
There are $O(\log n)$ adders on each path from an input to an output.
The tree thus has size  $O(n^2)$ and depth $O\((\log n)^2\)$ in total.)
For each bit of the binary input, we feed the \Zr{} of the appropriate power of two into one of the adders if that bit is a $1$, and zero into that adder if that bit is a $0$.
These additional networks have size $O(n^2)$ in total and depth $O(1)$, so the bounds for the tree apply to the entire network.

To convert from \Z{} to binary is even easier.
So called ``carry-save'' adders allow the tree of binary adders to have size $O(n^2)$ and depth only $O(\log n)$ (see Wallace [W1]).
Thus the size and depth required for conversion are dominated by those required in the 
binary-to-\Z{} direction.

For multiplication, division and square-root extraction, there are binary algorithms whose size and depth requirements are also dominated by those for binary-to-\Z{} conversion.
(Wallace [W1] gives networks for multiplication of size $O(n^2)$ and depth $O(\log n)$ (again using trees of carry-save adders).
He also gives networks of size $O(n^2)$ and depth $O\((\log n)^2\)$ for division and square-root extraction.)

There are much more efficient (in terms of both size and depth) networks known for binary multiplication, division and square-root extraction (see Pippenger [P] for a survey).
It is a challenging open problem to try to match their efficiency with networks for the corresponding operations in \Zr, either by improving the conversion process that is the current bottleneck, or by
finding more efficient algorithms for these operation directly in \Zr.
\sk

\heading{5. Acknowledgment}

The research reported in this paper was supported in part by grant CCF 0917026 from the National Science Foundation.
\sk

\heading{6. References}

\ref A; A. Apostolico and A. S. Fraenkel;
``Robust Transmission of Unbounded Strings Using Fibonacci Representations'';
IEEE Trans.\ Inform.\ Theory; 33:2 (1987) 238--245.

\ref F1; P. Fenwick;
``\Z{} Integer Arithmetic'';
Fibonacci Quart.; 41:5 (2003) 405--413.

\ref F2; A. S. Fraenkel and S. T. Klein;
``Robust Universal Complete Codes for Transmission and Compression'';
Discr.\ Appl.\ Math.; 64 (1996) 31--55.

\ref F3; H. T. Freitag and G. M. Phillips;
``On the \Z{} Form of $F_{kn}/F_n$'';
Fibonacci Quart.; 34:5 (1996) 444--446.

\refinbook F4; H. T. Freitag and G. M. Phillips;
``Elements of \Z{} Arithmetic'';
in: G.~E. Bergum, A.~N. Philippou and A.~F. Horadam (Ed's);
Applications of Fibonacci Numbers;
v.~7, Kluwer Academic Publishers, 1998, pp.~129--132.

\ref L; P. Ligomenides and R. Newcomb;
``Multilevel Fibonacci Conversion and Addition;
Fibonacci Quart.; 22:3 (1984) 196--203.

\ref O; Yu. P. Ofman; 
``On the Algorithmic Complexity of Discrete Functions'';
Sov.\ Phys.\ Dokl.; 7 (1963) 589--591.

\ref P; N. Pippenger;
``The Complexity of Computations by Networks'';
IBM J. Res.\ Dev.; 31 (1987) 1032--1038.

\ref S; R. Silber;
``Wythoff's Nim and Fibonacci Representations'';
Fibonacci Quart.; 15:1 (1977) 85--88.

\ref T; G. J. Tee;
``Russian Peasant Multiplication and Egyptian Division in \Z{} Arithmetic'';
Austral.\  Math.\ Soc.\ Gaz.; 30:5 (2003) 267--276.

\ref W1; C. S. Wallace;
``A Suggestion for a Fast Multiplier'';
IEEE Trans.\ Computers; 13 (1964) 14--17.

\refbook W2; I. Wegener;
The Complexity of Boolean Functions;
John Wiley \& Sons Limited, and B.~G. Teubner, 1987.

\ref W3; W. A. Wythoff;
``A Modification of the Game of Nim'';
Nieuw Archief voor Wiskunde (2); 7 (1907) 199--202.

\ref Z; \'{E}. Zeckendorf;
``Repr\'{e}sentation des nombres naturels par une somme de nombres de
Fibonacci ou de nombres de Lucas'';
Bull.\ Soc. Roy.\ Sci.\ Li\`{e}ge; 41 (1972) 179--182.

\bye